\documentclass{iopart}
\usepackage{graphicx}
\graphicspath{{./},{/home/user/fig/publi/muon/}}

\begin{document}

\title{A novel type of splayed ferromagnetic order observed in Yb$_2$Ti$_2$O$_7$} 
\author{A. Yaouanc$^{1,2}$, P. Dalmas de R\'eotier$^{1,2}$, L. Keller$^3$, B. Roessli$^3$, A. Forget$^4$}
\address{$^1$ Universit\'e Grenoble Alpes, INAC-PHELIQS, F-38000 Grenoble, France}
\address{$^2$ CEA, INAC-PHELIQS, F-38000 Grenoble, France}
\address{$^3$ Laboratory for Neutron Scattering and Imaging, Paul Scherrer Institut, 5232 Villigen-PSI, Switzerland}
\address{$^4$ SPEC, CEA, CNRS, Universit\'e Paris-Saclay, CEA Saclay, 91191 Gif-sur-Yvette Cedex, France}

\date{\today}

\begin{abstract} 
The pyrochlore insulator Yb$_2$Ti$_2$O$_7$ has attracted the attention of experimentalists and theoreticians alike for about 15 years. Conflicting neutron diffraction data on the possible existence of magnetic Bragg reflections at low temperature have been published. Here we report the observation of magnetic Bragg reflections by neutron powder diffraction at 60~mK. The magnetic diffraction pattern is analyzed using representation theory. We find Yb$_2$Ti$_2$O$_7$ to be a splayed ferromagnet as reported for Yb$_2$Sn$_2$O$_7$, a sibling compound with also dominating ferromagnetic interactions as inferred from the positive Curie-Weiss temperature. However, the configuration of the magnetic moment components perpendicular to the easy axis is of the all-in--all-out type in Yb$_2$Ti$_2$O$_7$ while it is two-in--two-out in Yb$_2$Sn$_2$O$_7$. An overall experimental picture of the  magnetic properties emerges.
\end{abstract}

\pacs{75.25.-j,75.10.Jm,75.30.Kz,75.40.Gb}

\maketitle

{\sl Introduction} ---
The broad interest in geometrically frustrated magnetic materials such as pyrochlore insulator compounds $R_2M_2$O$_7$ ($Fd{\bar 3}m$ space group), where $R$ is a rare earth ion and $M$ a non magnetic element, arises from the large diversity of magnetic ground states and exotic fluctuations and excitations encountered \cite{Moessner06,Gardner10,Balents10,Lacroix11,Gingras14}. This is due to the combined effects of the geometrical constraint resulting from the topology of the pyrochlore lattice and the action of the different possible interactions. As a consequence, classical and quantum fluctuations are particularly strong. Among the interactions of interest, the crystal field at the $R$ position, the Heisenberg nearest-neighbour exchange interaction as well as its extension beyond the nearest-neighbour $R$ ions \cite{Wills06,Yavorskii08}, the anisotropy of the exchange interactions \cite{Curnoe07a,Curnoe08} and the inevitable dipolar interactions have been considered.

The most remarkable discovery in the realm of geometrically frustrated magnetic materials has been the spin-ice ground state of Ho$_2$Ti$_2$O$_7$ in 1997 \cite{Harris97}. It is characterized by dipolar correlations emerging from the topological spin-ice constraint. This translates into pinch-points in the neutron diffuse scattering patterns \cite{Fennell09} and magnetic monopole excitations \cite{Ryzhkin05,Castelnovo08,Morris09}. 

Because of its large ground-state degeneracy, a spin-ice compound does not order magnetically, i.e., it is in a so-called spin liquid state where the spins are strongly correlated although they do not display long range order. Quantum fluctuations are not expected to play a significant role for Ho$_2$Ti$_2$O$_7$. However, for some pyrochlore materials strong quantum transverse fluctuations \cite{Curnoe07a,Curnoe08} could stabilize new magnetic states such as a quantum spin liquid for which new exotic excitations are expected \cite{Balents10,Onoda10,Savary12,Shannon12,Gingras14}. 

One of the favourite candidates for this exotic ground state is Yb$_2$Ti$_2$O$_7$. Indeed, the exchange interactions are highly anisotropic, with a strong ferromagnetic component akin to the Ising exchange of spin ice \cite{Ross11}. The dominating ferromagnetic component is also suggested by the positive Curie-Weiss temperature \cite{Hodges01}. Remarkably, the Ising character of the exchange interaction is in contrast to the marked planar anisotropy of the Yb$^{3+}$ ground state Kramers doublet \cite{Hodges01,Bertin12}. The measured entropy provides evidence for a ground state doublet well isolated from the excited doublets \cite{Blote69,Hodges02,Chang14}.

The strong specific heat anomaly early reported by Bl\"ote and collaborators at $T_{\rm c} \simeq 0.21$~K for a polycrystalline sample \cite{Blote69} does not necessarily correspond to dipolar magnetic ordering. A hidden, i.e.\ non-dipolar, order has been invoked \cite{Thompson11a,DOrtenzio13}. Alternatively the coherent magnetic order could be so disrupted by crystal defects that any magnetic Bragg reflections would be too broad to be observable \cite{Ross11a}. Due to the long-standing controversy as to the presence of magnetic Bragg reflections \cite{Hodges02,Yasui03,Bonville04a,Gardner04,Ross09,Chang12,Robert15}, the magnetic nature of the ground state of Yb$_2$Ti$_2$O$_7$ at low temperature is still uncertain. Magnetic Bragg reflections are reported only in Refs.~\cite{Yasui03,Chang12} which describe data measured on the same crystal. The diversity of the specific heat \cite{Dalmas06a,Yaouanc11c,Chang12,Ross12} and muon spin relaxation ($\mu$SR) results \cite{Hodges02,Yaouanc03,DOrtenzio13,Chang14} adds to the difficulty in understanding the ground state. In addition to the sample quality issue first noticed in Ref.~\cite{Yaouanc11c}, the low temperature required for the observation of neutron magnetic Bragg reflections is an additional issue. This is particularly true for a powder sample as illustrated by the sibling compound Yb$_2$Sn$_2$O$_7$ for which only two out of the three powder neutron diffraction studies have reported the existence of magnetic reflections \cite{Yaouanc13,Lago14,Dun13}.

Our previous successful work for Yb$_2$Sn$_2$O$_7$ \cite{Yaouanc13} has motivated us to revisit the case of Yb$_2$Ti$_2$O$_7$. Here we report the observation of magnetic Bragg reflections at 60~mK by powder neutron diffraction for the titanate and the determination of its magnetic structure. The result of the analysis is based on crystal symmetry and is at variance with the previously proposed nearly collinear ferromagnetic structure \cite{Yasui03,Chang12}. 

{\sl Experimental} ---
The syntheses of the present sample and the one studied in Ref.~\cite{Hodges02} were similar. A single phase polycrystalline sample was prepared by heating the constituent oxides up to 1400$^\circ$C with four intermediate grindings. Interestingly, a similar procedure was used for Nd$_2$Sn$_2$O$_7$ and gave an excellent sample \cite{Bertin15}. A sharp singularity appears in the specific heat at $T_{\rm c} \approx 255$~mK for a previously prepared Yb$_2$Ti$_2$O$_7$ sample \cite{Dalmas06a}. Similar peaks have always been reported in the literature for powders of Yb$_2$Ti$_2$O$_7$ \cite{Blote69,Ross11a,DOrtenzio13,Chang14}.

The neutron diffraction measurements were performed at the cold neutron powder diffractometer DMC of the SINQ facility, Paul Scherrer Institut (PSI), Switzerland. In the evening before the experiment $\approx 10~{\rm g}$ of powder were inserted into a copper container filled at room temperature with 10~bars of helium and sealed. The container was mounted in a liquid helium cryostat. The dilution refrigerator was started the next morning and the neutron scattering experiment was started after stabilization of the temperature at 60~mK. The measurement lasted for almost two days at this temperature, followed by two and a half days at 1.5~K and again at 60~mK for another night. As just explained, two measurements were carried out at 60~mK, before and after the 1.5~K measurement. The appearance of additional intensity at 60~mK was reproducible, for the second run with lower counting statistics, of course, but clearly visible. The two data sets recorded at 60~mK were finally combined. The magnetic diffraction pattern results from the difference of the data sets at 60~mK and 1.5~K.

\begin{figure}
\begin{center}
    \includegraphics[width=0.75\textwidth]{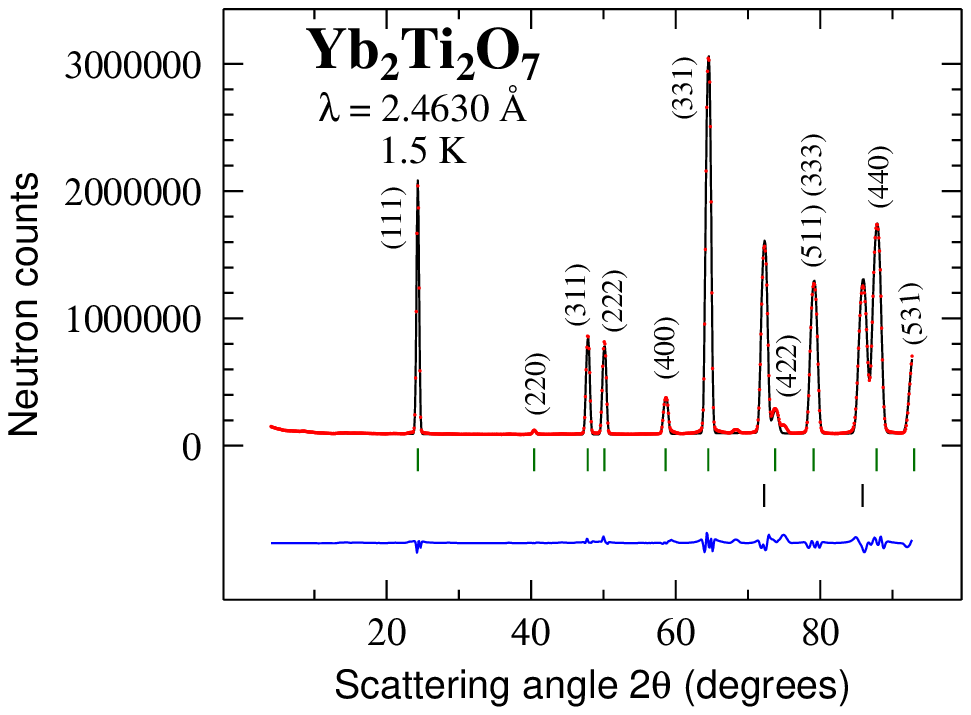}
\end{center}
\caption{(Color online) 
Neutron diffraction pattern for a powder of Yb$_2$Ti$_2$O$_7$ at 1.5~K. The intensity is plotted versus the scattering angle $2 \theta$. Neutrons of wavelength 2.4630~{\AA} were used. The solid black line results from a Rietveld refinement  \cite{Rietveld69}. The difference between the experimental data and the refinement is shown at the bottom. The observed reflections are labeled by Miller indices. The vertical markers indicate the positions of the Bragg peaks in the face-centered cubic $Fd\bar{3}m$ space group. The other two markers correspond to Bragg peaks of the Cu sample container. We get the lattice parameter of 
Yb$_2$Ti$_2$O$_7$, $a$ = 10.0220\,(5)~{\AA} and the free parameter for the oxygen position at the $2mm$ site, $x$ = 0.332\,(1). 
}
\label{neutron_nuclear}
\end{figure}

{\sl Crystal structure from neutron diffraction} ---
The diffraction pattern recorded at 1.5~K is shown in Fig.~\ref{neutron_nuclear}. 
Our results are consistent with available neutron \cite{Ross12} or synchrotron X-ray diffraction data \cite{Baroudi15} recognizing that these latter experiments have been performed at room temperature. Separate measurements at 1.9 and 20~K using a high resolution neutron powder diffractometer were performed\footnote{A detailed report of this structural study will be published elsewhere.}. The data confirm the high quality of the powder, with the presence of a ultra-minority rutile-type TiO$_2$ phase amounting to 0.9\%~wt. Its signature is visible at $2\theta \approx 68^\circ$ in Fig.~\ref{neutron_nuclear}. Concerning the pyrochlore phase, the partial occupancy of the Ti$^{4+}$ site by Yb$^{3+}$ ions --- the so-called stuffing already reported for crystals \cite{Ross12} --- is found to be 0.002\,(1), i.e. fully negligible.

\begin{figure}
\begin{center}
  \begin{picture}(281,219)
    \put(67,45){\large (a)}
    \put(0,0){\includegraphics[width=0.75\textwidth]{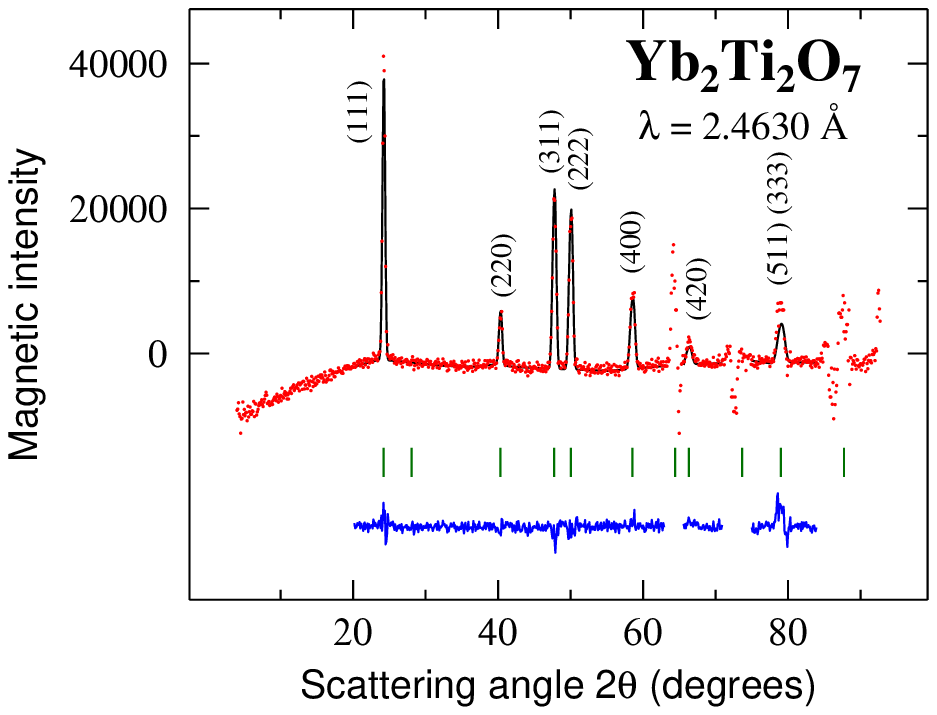}}
  \end{picture}
  \begin{picture}(281,219)
    \put(67,45){\large (b)}
    \put(0,0){\includegraphics[width=0.75\textwidth]{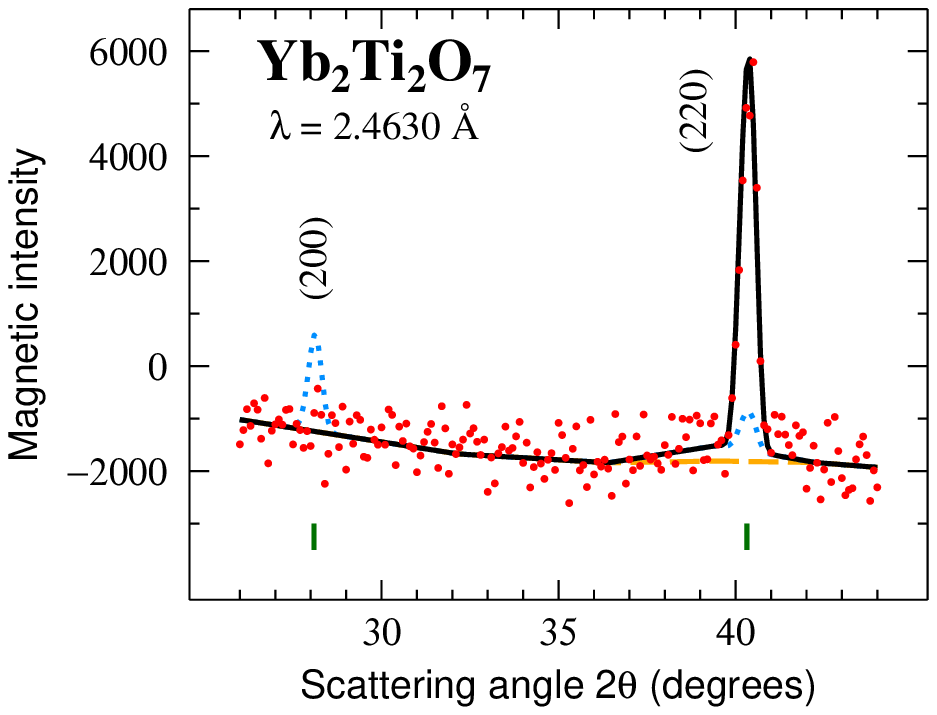}}
  \end{picture}
\end{center}
\caption{(Color online) (a) Magnetic neutron scattering pattern given by the difference between 60~mK and 1.5~K data sets for a powder of Yb$_2$Ti$_2$O$_7$ versus the scattering angle $2 \theta$. The experimental data are drawn as red dots.
  The observed magnetic reflections are labeled by Miller indices. The solid black line results from a Rietveld refinement according to the all-in--all-out splayed ferromagnetic model. The difference between the experimental data and the refinement is shown at the bottom. The vertical markers indicate the positions of the possible Bragg peaks in the ${\bf k}$ = 0 magnetic structure. Data near $2 \theta$ = 64$^\circ$ and 72$^\circ$ and below 20$^\circ$ and above 84$^\circ$ are not included in the refinement because they are strongly influenced by the sample environment.
  (b) Details of the data in the 24 -- 46 degrees angular range, showing a comparison of the prediction of the models considered in the main text. The black line is reproduced from panel (a) while the orange dashed line (light-blue dotted) line corresponds to the prediction for the collinear (two-in--two-out splayed) ferromagnetic structure. 
}
\label{neutron_result}
\end{figure}

{\sl Magnetic structure from neutron diffraction} ---
The magnetic diffraction pattern is displayed in Fig.~\ref{neutron_result}(a). Remarkably, 
magnetic reflections are observed at positions corresponding to the face-centered cubic reciprocal lattice. Therefore a magnetic order is found characterized by a ${\bf k}$ = 0 propagation wavevector. Since the width of the magnetic reflections is resolution limited, the magnetic order is long range.
Rietveld refinements of the magnetic diffraction pattern were attempted with FullProf \cite{Rodriguez93} for the magnetic structures allowed by the $Fd{\bar 3}m$ space group. As for Yb$_2$Sn$_2$O$_7$ \cite{Yaouanc13,Lago14} the data were found incompatible with the $Fd{\bar 3}m$ crystal symmetry.
Then we searched for a solution in the tetragonal space group $I4_1/amd$ with the help of the program BasIreps of the FullProf suite \cite{FullProf}. Among the highest non-isomorphic subgroups of $Fd{\bar 3}m$, this group is the only group simultaneously allowing ferromagnetic and antiferromagnetic components \cite{Mirebeau05}. In contrast to the case of Yb$_2$Sn$_2$O$_7$, no refinement according to this magnetic structure can account for the magnetic pattern of Yb$_2$Ti$_2$O$_7$; see the discussion section for more details. In fact, we found the magnetic pattern to be consistent with the magnetic structure described by spontaneous magnetic moment components as given in Table~\ref{table}. 
\begin{table}
\begin{center}
\begin{tabular}{ccccccc}
\hline
\hline
Site & $X$ & $Y$ & $Z$ & $m_{\rm sp}^X$ & $m_{\rm sp}^Y$ & $m_{\rm sp}^Z$\\
\hline
1 & $\frac{1}{2}$ & $\frac{1}{2}$ & $\frac{1}{2}$ & $\hphantom{-}0.30\,(1)$ & $\hphantom{-}0.30\,(1)$ & $0.86\,(1)$ \\
2 & $\frac{1}{2}$ & $\frac{1}{4}$ & $\frac{1}{4}$ & $\hphantom{-}0.30\,(1)$ & $-0.30\,(1)$ & $0.86\,(1)$ \\
3 & $\frac{1}{4}$ & $\frac{1}{2}$ & $\frac{1}{4}$ & $-0.30\,(1)$ & $\hphantom{-}0.30\,(1)$ & $0.86\,(1)$ \\
4 & $\frac{1}{4}$ & $\frac{1}{4}$ & $\frac{1}{2}$ & $-0.30\,(1)$ & $-0.30\,(1)$ & $0.86\,(1)$ \\
\hline
\hline
\end{tabular}
\end{center}
\caption{Components of the four spontaneous magnetic moments of the Yb$^{3+}$ ions in a tetrahedron. The site positions $X$, $Y$ and $Z$ and the moment components $m_{\rm sp}^X$, $m_{\rm sp}^Y$, $m_{\rm sp}^Z$, given in Bohr magneton units at the temperature of 60~mK, are expressed respective to the cubic axes. The uncertainties are of statistical origin.}
\label{table}
\end{table}
Relative to the stannate, the signs of the $X$ and $Y$ components for sites $1$ and $4$ are reversed.

Looking for the symmetry elements that keep the magnetic structure invariant, the $Im'm'a$ group was inferred to be the possible magnetic space group in which Yb$_2$Ti$_2$O$_7$ orders \cite{note_1}. The prime superscript indicates that the symmetry operation is combined with time reversal. The corresponding structural space group is $Imma$. Interestingly, this latter space group
is one of the subgroups of $I4_1/amd$ \cite{Hahn05}. The $a$ and $b$ axes of the orthorhombic cell are shorter than the cubic axes by a factor $\sqrt{2}$ and are rotated by 45$^\circ$ relative to them.
A signature of the displacement of the atoms from their position in the cubic space group would be a splitting or a broadening of the Bragg peaks when crossing the magnetic transition. An alternative sensitive signature of such a displacement would be the appearance of nuclear scattering intensity at a position forbidden in the $Fd{\bar 3}m$ space group but authorised in the $Imma$ space group, e.g.\ (002). Since neither of the effects is observed, the structural changes are negligible. 
The 16d site of $Fd{\bar 3}m$ occupied by the Yb$^{3+}$ ions transforms and splits into two sites, namely 4b and 4c, in $Imma$. In terms of symmetry the moments at the two sites are allowed to differ. However, within our experimental uncertainties we found them equal, consistent with the $^{170}$Yb M\"ossbauer spectroscopy results \cite{Hodges02}. Their amplitude is $m_{\rm sp}$ = 0.95\,(2)\,$\mu_{\rm B}$, a value reasonably close to the estimate $1.1 \, (1) \,\mu_{\rm B}$ from neutron diffraction on a single crystal \cite{Yasui03}, although interpreted with a different magnetic structure --- see below ---  and 1.15~$\mu_{\rm B}$ from M\"ossbauer spectroscopy \cite{Hodges02}.

\begin{figure}
\begin{center}
\includegraphics[width=0.6\textwidth]{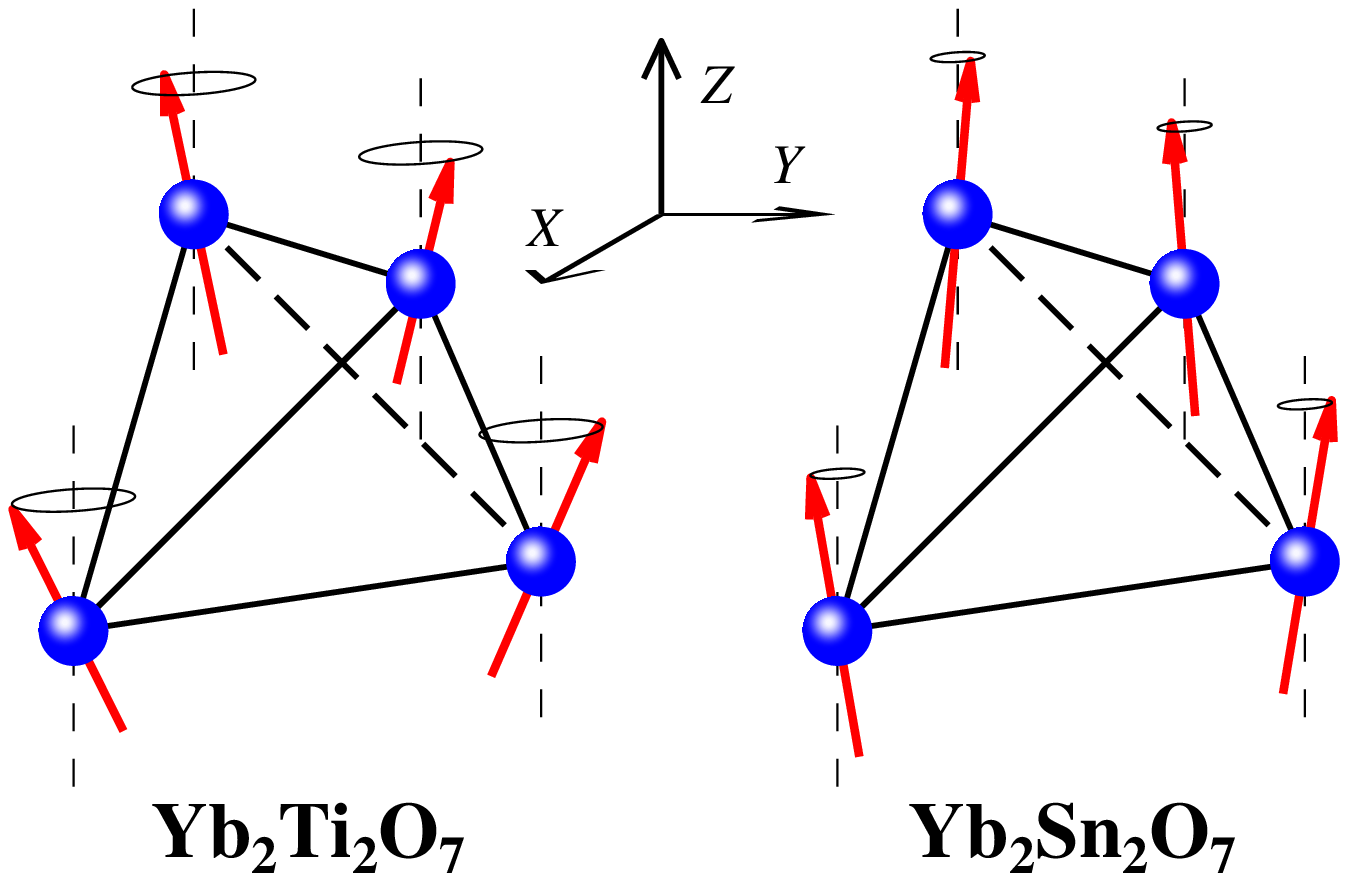}\\
\includegraphics[width=0.6\textwidth]{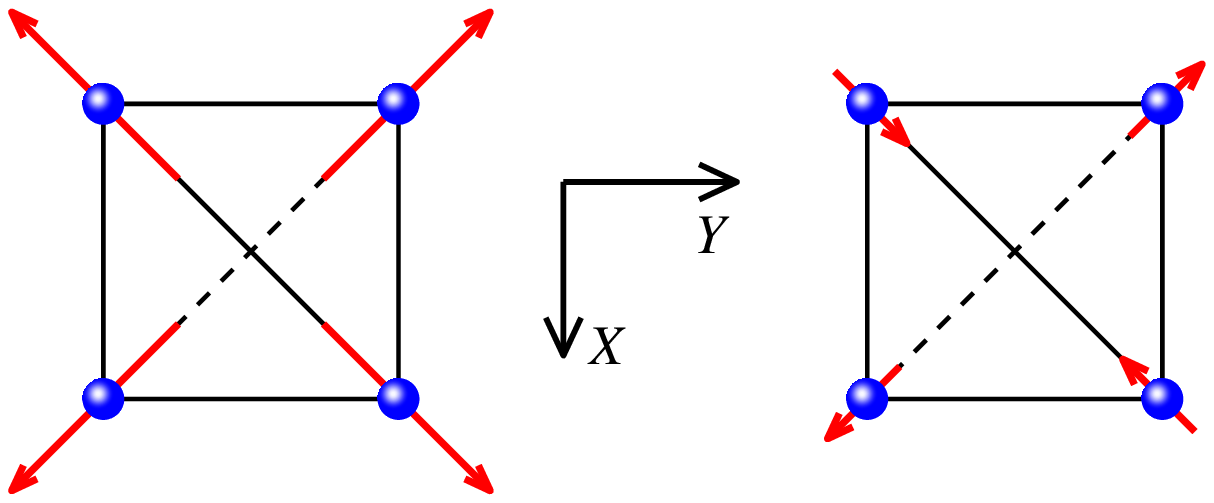}
\end{center}
\caption{(Color online)
Comparison of the magnetic structures inferred from neutron powder diffraction for Yb$_2$Ti$_2$O$_7$ and Yb$_2$Sn$_2$O$_7$.  Reference frame ($X, Y, Z$) corresponds to the cubic axes. (Upper panel) Perspective drawing of the magnetic moments (arrows) of the Yb$^{3+}$ ions (spheres) of a tetrahedron for the two compounds. (Lower panel) Projection of the tetrahedon magnetic moments in the ($X,Y$) plane. In Yb$_2$Ti$_2$O$_7$, while all the projected components point outwards of the represented tetrahedron, in the neighbouring tetrahedra they point inwards. In Yb$_2$Sn$_2$O$_7$, for all the tetrahedra, two components are directed inwards and the other two outwards. For these reasons the Yb$_2$Ti$_2$O$_7$ and Yb$_2$Sn$_2$O$_7$ arrangements of the moment component perpendicular to the ferromagnetic axis are respectively referred as all-in--all-out and two-in--two-out.
}
\label{mag_structure}
\end{figure}

In Fig.~\ref{mag_structure} we illustrate the magnetic structures for the two ytterbium compounds.
While the projections of the magnetic moments into the $XY$ plane normal are in an all-in--all-out configuration for the titanate, they adopt a two-in--two-out configuration for the stannate.
The angle of a magnetic moment relative to the $Z$ cube edge is $\phi \approx 26^\circ$. While $\phi$ is clearly larger than for Yb$_2$Sn$_2$O$_7$ ($\phi \approx 10^\circ$ \cite{Yaouanc13}), $m_{\rm sp}^Z$ is still predominant relative to $m_{\rm sp}^\perp = \sqrt{2} m_{\rm sp}^X$. Hence Yb$_2$Ti$_2$O$_7$ belongs to the class of the splayed ferromagnets as its sibling  Yb$_2$Sn$_2$O$_7$. This is consistent with their positive Curie-Weiss temperature \cite{Hodges01,Yaouanc03} and the single-crystal magnetization of Yb$_2$Ti$_2$O$_7$ \cite{Lhotel14a}. Although strictly speaking, the three-fold symmetry axis is no longer present in the orthorhombic structure, since the deformation is very small, it is of interest to compute the angle between this direction and the magnetic moment. We find 29\,(1)$^\circ$ for sites 1 and 4 and 81\,(1)$^\circ$ for sites 2 and 3 \footnote{While, in Yb$_2$Sn$_2$O$_7$, a single angle of 65$^\circ$ between the magnetic moment and the local three-fold axis was deduced from M{\"o}ssbauer spectroscopy, this angle cannot be directly measured in Yb$_2$Ti$_2$O$_7$. This is due to a vanishing quadrupolar interaction in the latter compound \cite{Hodges02}.}.
The site numbering refers to Table~\ref{table}.

{\sl Discussion} ---
The collinear ferromagnetic structure \cite{Yasui03} previously deduced from measurements on a single crystal with a specific heat anomaly at $T_{\rm c} \approx 195$~mK \cite{Chang12} is inconsistent with our result. In Table~\ref{comparison} we compare the magnetic intensities measured at the position of the first three Bragg peaks authorised in a face-centred cubic structure with the intensities obtained for the best fit to the collinear ferromagnetic structure. For reference we also compare the best fits to the two-in--two-out and to all-in--all-out models. Figure~\ref{neutron_result}(b) graphically illustrates the comparison. The last model is obviously the only one consistent with the data.
\begin{table}
  \caption{\label{comparison} Comparison of three models for the magnetic structure refinement. The rows give the integrated intensity in arbitrary units at three Bragg positions for the best fit to three models as described in the main text. The last row corresponds to the experimental result. For each of the three models the components for the magnetic moment and its modulus are given. Finally, as a numerical indicator of the fitting agreement, the weighted profile factor \cite{Rodriguez93} $R_{\rm wp}$ is reported. For reference, its expected value is 4.40.}
  \center
  \begin{tabular}{llllllll}
    \hline
                           & \multicolumn{3}{c}{Bragg peaks} & $m_{\rm sp}^X$& $m_{\rm sp}^Z$ & $m_{\rm sp}$ & $R_{\rm wp}$\cr
                           & \multicolumn{1}{c}{(111)} & \multicolumn{1}{c}{(200)} & \multicolumn{1}{c}{(220)} & ($\mu_{\rm B}$) & ($\mu_{\rm B}$) & ($\mu_{\rm B}$) & \cr
Collinear FM               & 18318 & \hphantom{00}0 & \hphantom{000}0 & 0 & 0.90\,(2) & 0.90\,(2) & 9.31\cr
Two-in--two-out splayed FM & 22031 & 873 & \hphantom{0}410 & 0.14\,(2) & 0.85\,(1) & 0.88\,(2) & 8.91\cr
All-in--all-out splayed FM & 18591 & \hphantom{00}0 & 4129 & 0.30\,(1) & 0.86\,(1) & 0.95\,(2) & 5.33 \cr
Experiment                 & 19535\,(480) & $\le 155$ & 4117\,(152)\cr
    \hline
    \hline
  \end{tabular}
\end{table}

As already mentioned, the size of the magnetic moment of the Yb$^{3+}$ ion inferred from our neutron data is reasonably consistent with the expectation from M\"ossbauer spectroscopy \cite{Hodges02}. The residual difference could be tentatively assigned to short-range magnetic correlations observed in several instances \cite{Chang12,Maisuradze15,Robert15}, while the diffraction experiment reported here only picks up the long-range ordered component. The signature of these short-range correlations seems to disappear below $T_{\rm c}$ \cite{Chang12}, but they are still observed in $\mu$SR data \cite{Maisuradze15}.

A large body of experimental results on Yb$_2$Ti$_2$O$_7$ is available in the literature. Because of the mentioned sample quality issues, to get reliable characterizations a selection might be necessary. We propose the cautious criterion of considering only results obtained on polycrystalline samples and on crystals for which magnetic Bragg reflections have been observed. When present in crystals, a sharp peak in the specific heat is observed in the range 0.20 -- 0.25~K as for polycrystals.

The thermal variation of the entropy deduced from specific heat data is consistent with a well isolated Yb$^{3+}$ crystal-field Kramers doublet \cite{Hodges02,Chang14}. Apart from the magnetic order described here, measurements on polycrystalline samples have revealed a sharp change in the spin dynamics at $T_{\rm c}$ in zero field \cite{Yaouanc03}. Signatures of short-range magnetic correlations both in the paramagnetic state through a specific-heat measurement \cite{Yaouanc11c} and in the ordered state from $\mu$SR data \cite{Yaouanc13a,Dalmas14} have been found. Short-range correlations in the ordered state were certainly unexpected. In terms of characteristic fluctuation times, there is a large range of magnetic fluctuation modes, some of them being almost similar in the paramagnetic (still below $\approx 0.7$~K) and ordered phases \cite{Maisuradze15,Robert15}. The magnetization curves along the [100] and [110] directions have been determined \cite{Lhotel14a}. Polarized neutron scattering on crystals shows the presence of a diffuse [111]-rod of scattering and pinch-point features which develop on cooling \cite{Chang12}. These features are suppressed at $T_{\rm c}$. The magnetic susceptibility below $T_{\rm c}$ is stronger in the titanate than in the stannate \cite{Maisuradze15}.  

A possible theoretical framework for understanding this set of properties could be to consider the paramagnetic phase as the manifestation of a magnetic Coulomb phase with strong quantum fluctuations. It is then tempting to attribute the transition at $T_{\rm c}$ to the condensation of magnetic monopoles through Higgs's mechanism \cite{Chang12}. However, much work is needed to support this picture. Since spin dynamics is exotic, an inelastic neutron scattering study on reliable crystals is necessary. In particular it should be determined whether a gap exists in the excitation spectrum. A scattering technique signature of molecular spin structures such as spin loops involved in the so-called persistent spin dynamics \cite{Yaouanc15} would be of great help. The fact that no spontaneous field has been detected by the $\mu$SR technique \cite{Hodges02,DOrtenzio13,Chang14} is a proof of the existence of pecular spin dynamics related to the Coulomb field \cite{Bertin15} and probably the short-range correlations.

On the theoretical side, the ubiquitous slow spin dynamics \cite{Hodges02,Yaouanc03,Yaouanc13a,Lhotel14a,Maisuradze15}, observed even below $T_{\rm c}$ has to be described. In fact, the most amazing property is not the presence of magnetic order at low temperature --- conventional magnetic compounds do display magnetic ordering at low temperature --- but the existence of exotic spin correlations in both the paramagnetic and ordered regimes, with the possibility of competing mechanisms at their origin \cite{Jaubert15}. 

{\sl Conclusions} ---
Yb$_2$Ti$_2$O$_7$ does show magnetic Bragg reflections at low temperature. The structure adopted by its magnetic moments is of the all-in--all-out splayed ferromagnetic type. The actual magnetic moment arrangement --- different from the one found for Yb$_2$Sn$_2$O$_7$ --- should give constraints to the parameters of a Hamiltonian describing the system. While the compound is definitively not a spin liquid system at low temperature, it displays quite exotic spin correlation properties. Although we suspect that they stem from the geometrical constraint imposed by the pyrochlore lattice, this proposal needs to be confirmed by further experimental and theoretical works. 

We acknowledge H. Grimmer for an enlighting discussion. This work is based on experiments performed at the Swiss spallation neutron source SINQ, Paul Scherrer Institut, Villigen, Switzerland.

{\sl Note added} ---
Two papers on Yb$_2$Ti$_2$O$_7$ have recently been published. In one of them, magnetic Bragg reflections have also been observed for a powder sample \cite{Gaudet16}. The inferred magnetic structure is different from the structure derived in the present report. We note that the limited statistical accuracy of the data reported in Ref.~\cite{Gaudet16} makes it difficult to decide between the two models: compare, e.g.\ the prediction for the intensity at Bragg peak (220) (Table~\ref{comparison}) and Fig.~2(a) of Ref.~\cite{Gaudet16}. Concerning the detection reported by Gaudet {\em et al} of magnetic intensity above the temperature of the heat capacity peak, we note that the magnetic transition is first order and that the magnetic and paramagnetic phases were found to coexist over a sizable temperature interval \cite{Hodges02}. We suggest this is an explanation for the persistence of a magnetic signal above $\approx 0.26$~K in Ref.~\cite{Gaudet16}. The work by Gaudet {\em et al} also shows that the magnetic excitation spectrum is a gapless continuum. The same group has also reported on crystalline electric field (CEF) measurements and analysis for a stoichiometric and a stuffed sample \cite{Gaudet15}. The CEF level scheme is similar in both cases and agrees with the previously published scheme \cite{Bertin12}.

\section*{References}
\bibliography{Yb2Ti2O7_dmc_iop.bib}

\begin{thebibliography}{10}
\bibitem{note_1}
For an introduction to magnetic space groups, see, e.g. Ref.~\cite{Rodriguez12}
  and references therein. Note that the magnetic structure symmetry elements
  must be consistent with the underlying crystal lattice irrespective of the
  first order character of the magnetic transition \cite{Bertaut81}.
\end{thebibliography}
\bibliographystyle{iopart-num}

\end{document}